# Time-zero Efficiency of European Power Derivatives Markets

Juan Ignacio Peña[a] and Rosa Rodriguez[b]

**Abstract**

We study time-zero efficiency of electricity derivatives markets. By time-zero efficiency is meant a sequence of prices of derivatives contracts having the same underlying asset but different times to maturity which implies that prices comply with a set of efficiency conditions that prevent profitable time-zero arbitrage opportunities. We investigate whether statistical tests, based on the law of one price, and trading rules, based on price differentials and no-arbitrage violations, are useful for assessing time-zero efficiency. We apply tests and trading rules to daily data of three European power markets: Germany, France and Spain. In the case of the German market, after considering liquidity availability and transaction costs, results are not inconsistent with time-zero efficiency. However, in the case of the French and Spanish markets, limitations in liquidity and representativeness are challenges that prevent definite conclusions. Liquidity in French and Spanish markets should improve by using pricing and marketing incentives. These incentives should attract more participants into the electricity derivatives exchanges and should encourage them to settle OTC trades in clearinghouses. Publication of statistics on prices, volumes and open interest per type of participant should be promoted.

[a] Corresponding author. Universidad Carlos III de Madrid, Department of Business Administration, c/ Madrid 126, 28903 Getafe (Madrid, Spain). ypenya@eco.uc3m.es; [b] Universidad Carlos III de Madrid, Department of Business Administration, c/ Madrid 126, 28903 Getafe (Madrid, Spain). rosa.rodriguez@uc3m.es.

# Time-Zero Efficiency of European Power Derivatives Markets

## Abstract

We study time-zero efficiency of electricity derivatives markets. By time-zero efficiency is meant a sequence of prices of derivatives contracts having the same underlying asset but different times to maturity which implies that prices comply with a set of efficiency conditions that prevent profitable time-zero arbitrage opportunities. We investigate whether statistical tests, based on the law of one price, and trading rules, based on price differentials and no-arbitrage violations, are useful for assessing time-zero efficiency. We apply tests and trading rules to daily data of three European power markets: Germany, France and Spain. In the case of the German market, after considering liquidity availability and transaction costs, results are not inconsistent with time-zero efficiency. However, in the case of the French and Spanish markets, limitations in liquidity and representativeness are challenges that prevent definite conclusions. Liquidity in French and Spanish markets should improve by using pricing and marketing incentives. These incentives should attract more participants into the electricity derivatives exchanges and should encourage them to settle OTC trades in clearinghouses. Publication of statistics on prices, volumes and open interest per type of participant should be promoted.

## 1. Introduction

The ongoing liberalization process of electricity markets in Europe has many challenges, such as increasing efficiency in spot and derivatives markets, and promoting European market integration. The extent to which these challenges are met is of paramount importance for market participants, regulators and external investors who include power derivatives in their investment portfolios. Besides that, the European Commission (EC) is also promoting actions to ensure efficient, safe derivatives markets, including power derivatives markets (EC, 2009).

In this paper, we focus on one crucial, but hitherto neglected, aspect of informational efficiency of European power derivatives markets. Extant evidence largely concentrates on studying whether spot and (usually near-term) derivatives prices are linked by a set of efficiency conditions, related with size and characteristics of the risk premium. Key limitations of this literature are difficulties in capturing the complex behaviour of the forward risk premium across contracts and over time, see Cartea and Villaplana (2008) among others. Benth, et al. (2008) provide an explanation for sign and magnitude of the market risk premium by modelling market players and their risk preferences by applying certainty equivalence principles. However, given that ex-ante risk premium between spot and derivatives contracts is not directly observable from market data, it is always dependent on the spot price chosen[a].

An alternative approach is to study the degree of time-zero efficiency of electricity derivatives markets using derivatives contracts only. This is a minimal condition for market efficiency and, consequently, inferences regarding resource-allocation efficiency of these markets based on our results are limited. No-arbitrage condition between swap prices and their corresponding replicating portfolios are well known. However, as far as we know, in the case of European power derivatives markets, extant literature does not address this issue. In other words, it seems that this basic efficiency test has not been applied to those derivatives markets. Ours is a first step to fill this gap in the literature. Furthermore, our approach has the advantage of dispense altogether with the problematic estimation of risk premia.

---

[a] In the application of their model to the EEX market Benth, et al. (2008) employ a two-factor spot market model without spikes and with a constant level to which the prices mean-revert. However, Bierbrauer, et al. (2007) suggest that, in the case of the EEX market, a regime-switching model using two regimes and with a Gaussian distribution in the spike regime outperforms alternative models.

In this context, time-zero efficiency means a sequence of prices of the derivatives contracts having the same underlying asset but different times to maturity which implies that prices comply with a set of efficiency conditions that prevent profitable arbitrage opportunities. In this paper we investigate whether statistical tests, based on the law of one price, and trading rules, based on price differentials, are useful for assessing time-zero efficiency in a given power derivatives market.

Given that, by far, the most liquid derivatives contracts in the electricity markets are forwards, futures and swaps[b], we apply tests using swap prices data from three of the biggest electricity markets in the Euro area. The German Market, which is Europe's largest power market in terms of consumption, the French Market, which is the second largest European market, and the Spanish Market which is the fifth largest market (EC, 2014). In the case of the German and French markets, we use data from EEX (European Energy Exchange). In the case of the Spanish market, we use data from OMIP (Iberian Energy derivatives Exchange).

---

[b] In most electricity markets, forward and futures contracts guarantee delivery of the electricity over a period of time (e.g. monthly or yearly contracts) rather than at a fixed future time. As Benth and Koekebakker (2008) argue the nature of these contracts are very similar to a swap exchanging a fixed price for floating (spot) electricity price during the defined period. In fact, swap contracts are integrals of traditional fixed delivery time forward contracts. If the contract is purely financial, the contract is settled in cash against the system price, during the delivery period. Thus, financial electricity contracts are swap contracts, exchanging a floating spot price against a fixed price. On the other hand, if the contract is taken to physical delivery, it may be converted into what is the equivalent of a one-month swap.

Salient causes of problems in European electricity markets are market inefficiency, market power, inelastic demand, and constrained supply. Regarding the first point, extant literature largely focuses on assessing either (i) the degree of spatial integration across European spot markets or (ii) the degree of integration between spot prices and (near-term) derivatives prices in a given country. The degree of spatial integration across spot (or day-ahead) European markets is addressed in Bower (2002), Boiselau (2004), Zachman (2008), and Bunn and Gianfreda (2010) among others. The overall conclusions suggest that significant spatial integration exists, although this integration is not homogeneous across Europe. In addition, this integration tends to increase over time. Regarding the issue (ii) most papers focus on the extent to which, in a given market, forward prices are unbiased predictors of future spot prices or whether the basis (the difference between current spot and forward prices) contains useful information for forecasting changes in spot prices. See Lucia and Schwartz (2002), Wilkens and Wimschulte (2007), Cartea and Villaplana (2008) and Furió and Meneu (2010) among others. Bunn and Gianfreda (2010) report some results on the predictability of forward prices of the French, German, British, Dutch and Spanish power markets and they conclude that there is an efficiency problem in all these markets since optimality conditions are never satisfied simultaneously, and they observe perfect forecast ability only in one market. On the specific issue of arbitrage opportunities, Capitán Herráiz (2014) studies the extent to which they are present and are profitable in the case of OMIP auctions, CESUR auctions and VPP auctions[c] in the scope of MIBEL (the Iberian Electricity

---

[c] In CESUR auctions, the agents trade energy contracts for last resort supply (Contratos de Energía para el

market), by means of an analysis of the existence of arbitrage opportunities amongst trading mechanisms and maturities. Nevertheless, the important issue of whether, in a given European electricity market, prices of derivatives contracts with different time to maturity obey time-zero efficiency conditions that prevent profitable arbitrage opportunities has not been studied so far[d]. This paper aims to fill this gap in the literature.

To do so, we present several tests for assessing the time-zero efficiency of a swap power market. First, we test the law of one price between a swap contract with a given maturity, say one year, and its replicating portfolio, constructed using contracts with shorter maturities, say one month, by means of unit roots test, cointegration analysis, price convergence tests and price variance tests. Second, we apply some trading rules based on price differentials and on violations of zero-time arbitrage conditions. We apply tests and trading rules using eight years of daily data of prices of monthly, quarterly and yearly swap contracts from the three markets we study. The sample spans data of the German market from June 1, 2004 until December 31, 2012, of the French market from July 3, 2006 to March 19, 2014 and of the Spanish market from July 3, 2006 until June 6, 2014. We take into account relative liquidity of contracts and of their replicating portfolios, as well as transaction costs such as bid-ask spreads, trading fees and clearing fees.

In the case of the German market, after considering liquidity availability and transaction costs, results are largely consistent with time-zero efficiency. In the case of the French and Spanish markets, limitations in liquidity and representativeness are challenges that prevent definite conclusions. We recommend some policy measures to meet these challenges.

---

Suministro de Último Recurso) . In VPP auctions, the agents trade energy contracts for Virtual Power Plant capacity.
[d] Shahidehpour, Yamin and Li (2002) provide an overview of arbitrage strategies in electricity markets, but they focus largely on cross-commodity arbitrage.

Specifically, liquidity in French and Iberian markets should improve by using pricing and marketing incentives. These incentives should attract more participants into the electricity derivatives exchanges and should encourage them to settle OTC trades in clearinghouses. Publication of statistics on prices, volumes and open interest per type of participant should be encouraged. The results of this paper are therefore relevant for evaluating liberalized electricity markets in Europe.

The remainder of this paper is organized as follows. Section 2 presents time-zero efficiency tests. After we describe the data in Section 3, we report results of empirical analysis in Section 4. Section 5 discusses the impact of liquidity and representativeness. Section 6 contains conclusions and policy recommendations.

## 2. Time-zero efficiency tests

In this section, we present definitions for the variables we use as the basis for time-zero efficiency tests and present some guidelines for its practical implementation.

**2.1 Definitions: Swaps and replicating portfolios**

Assume T<∞ and let ($\Omega$,$F$,$Q$) be a complete filtered probability space, with an increasing and right-continuous filtration $\{F_t\}_{t \in [0,T]}$ where, as usually, $F_o$ contains all sets of probability zero in $F$. We assume the market trades swap contracts with different delivery periods and a bond that yields a constant risk free rate $r > 0$, so futures and forward prices are equal. Consider the price $F_i(t,T)$ of a swap contract with expiry date $T = 1, \ldots, N$, which is also the start of the delivery period, time to delivery (maturity) $\tau = (T - t)$, delivery length period given by the subscript $i = M, Q, Y$ (i.e.. $i$=M for monthly contracts, $i$=Q for quarterly contracts and *i*=Y for yearly contracts)[e]. All contracts are settled against the daily average spot price during the delivery period and, in agreement with market practice, we call them electricity swaps. Hence, $F_i(t,\boldsymbol{T})$ denotes the price vector of a completely observable swap curve at time *t*, where vector $\boldsymbol{T}$ indicates the different swap contracts starting delivery dates which are available at trading date *t*, and vector subscript $\boldsymbol{i}$ denotes the underlying asset (delivery period) over which is defined each curve's swap contract. Notice that, in a general term structure model, it is considered the evolution of the continuum of swap contracts for all the possible expirations. Since only subsets of them trade in the market, we test only for tradable contracts that are liquid enough. We concentrate in monthly, quarterly and yearly contracts because, on average, they are responsible of the 8%, 16% and 75% respectively of the total volume of futures in the German market, of the 9%, 21% and 70% in the French market and of the 17%, 61% and 21% in the Spanish market[f].

In what follows, we use the following notation to simplify the exposition. Regarding monthly contracts, we denote them Mj(t) where j=1,2,…,J. For example, if t is any day in January, M1(t) is the price, at time t, of the monthly swap contract that matures at the end of the current month (January) and provides delivery of electricity at a fixed price during next month (February). Regarding quarterly contracts, we denote them Qbc(t) where b=1,2,…,B and c=1,2,…,C. Notice that b refers to number of quarters ahead in which the contract delivers the electricity and c refers to quarter in which contract trades. For example, if t is any day in January, February, or March, Q11(t) refers to the price at time t (first quarter, c=1) of the quarterly swap contract for the following quarter (one quarter ahead, b=1), corresponding to April, May and June. If *t* is any day in April, May, or June (second quarter, c=2), Q12(t) refers to the price at time t of the quarterly swap contract for the next quarter, corresponding to July, August and September (one quarter ahead, b=1). Notice that although Q11 and Q12 both refer to next quarter, they are fundamentally different contracts because they apply to different seasons; Q11 applies to Spring and Q12 applies to Summer. Regarding yearly contracts, we denote them Yf(t) where f=1,2,...,F. For example, if t is any day in 2015, Y1(t) is the price at time t of the yearly swap contract for 2016. Replicating portfolios for quarterly and yearly contracts are built in such a way that they correspond to payoffs generated by a given swap contract. We denote these replicating portfolios as AQbc(t) for quarterly contracts and AYf(t) for yearly contracts.

---

[e] For example, the swap price $F_M(t,1)$ is the price, at time t, of the contract that matures at the end of the current month (e.g. January) and provides delivery of electricity at a fixed price during the next month (e.g. February). This contract is the M+1 or M1 monthly contract in market parlance. Similarly, $F_Q(t,2)$ is the price of the Q2 quarterly contract, and $F_Y(t,3)$ is the price of the Y3 yearly contract.

[f] In our sample, short term contracts (i.e. maturity lower than one month) are responsible of around 1%, 2% and 1% of total volume in the case of Germany, France and Spain respectively (EEX, 2013).

We define the replicating portfolio as the weighted average of the appropriate swap contracts with lower time to maturity than the original contract. Weights depend on the present value of each contract included in the replicating portfolio. For the sake of clarity, in what follows, we present the case of quarterly contracts; the case of yearly contract is similar. Let us introduce a discount function *w(j)*, with components $w(j) = (1+r)^{-j/12}$. We define the weight function as follows : $g(t,j) = \frac{w(j)}{\sum_j w(j)}$, indexed by set *j*={1,2,...,*J*} , corresponding to monthly contracts M*j*(t) included in the replicating portfolio. The weight function integrates to one. Therefore, replicating portfolio AQbc(t) is calculated as $AQbc(t) = \sum_j g(t,j)Mj(t)$. For instance, in the case of the quarterly contract Q11, in January, set *j* is *j* ={3,4,5} and replicating portfolio is AQ11 = w(t,3) M3(t) + w(t,4) M4(t) + w(t,5) M5(t). The elements of the weight function are w(t,3) = w(3)/[w(3)+w(4)+w(5)], w(t,4) = w(4)/[w(3)+w(4)+w(5)], and w(t,5) = w(5)/[w(3)+w(4)+w(5)]. For low values of *r* and *j,* the weight function can be approximated by $g(t,j) = |j|^{-1}$ where $|j|$ is the cardinality of the set $j$[g]. We use this approximation in what follows[h]. The general expression for replicating portfolios of quarterly contracts by using monthly contracts M is

$$AQbc = \frac{\sum_{i=0}^{2} M[j - 3c - b + K + i]}{3} \quad ; \quad j = 1, \ldots, J \,; b = 1, \ldots, B; \; c = 1, \ldots, C \quad (1)$$

Where $K=4b+2$ if j=1,4,…,J-2, $K=4b$ if j=2,5,…,J-1 and $K=4b - 2$ if j=3,6,…,J. Notice that the structure of replicating portfolios changes over time because it depends on the available contracts in each month. For instance, the replicating portfolio of Q11 is defined as AQ11 = (1/3)(M3+M4+M5) in January, as AQ11 = (1/3)(M2+M3+M4) in February and as AQ11 = (1/3)(M1+M2+M3) in March.

The general expression for replicating portfolios of yearly contracts AYf , using quarterly contracts Qbc is

$$AYf = \frac{\sum_{i=1}^{4} Q[b(i) * 4(f - 1), c]}{4} \quad ; \quad f = 1, \ldots, F; \; c = 1, \ldots, C \quad (2)$$

Where if c =1, then b(i) = c+3,…,c+6, if c =2, then b(i) = c+1,…,c+4, if c =3, then b(i) = c-1,…,c+2, and if c =4, then b(i) = c-3,…,c. Similarly to quarterly portfolios, the structure of these replicating portfolios changes over time because it depends on the available contracts in each month. For instance, the replicating portfolio of Y1 is defined as AY1 =

---

[g] We thank an anonymous referee for clarifying remarks on this point.
[h] In the case of the AQ11, the approximation implies that weights are w(3) = w(4) = w(5) = 1 and w(t,3) = w(t,4) = w(t,5) = 0.3333. As a comparison, assuming r = 2.2543% (which was the average 3-month EURIBOR during our sample 2004-2012) , exact weights are w(3) = 0.9943, w(4) = 0.9925, w(5) = 0.9906, and w(t,3) = 0.3339, w(t,4) = 0.3333, and w(t,5) = 0.3327.

(1/4)(Q41+Q51+Q61+Q71) in the first quarter (January, February and March), it is defined as AY1 = (1/4)(Q32+Q42+Q52+Q62) in the second quarter (April, May and June), as AY1 = (1/4)(Q23+Q33+Q43+Q53) in the third quarter (July, August and September) and as AY1 = (1/4)(Q14+Q24+Q34+Q44) in the fourth quarter (October, November and December).

**2.2 Statistical tests**

We perform four tests to gauge the time-zero efficiency of power derivatives markets. First, we evaluate the predictability of a swap price series and of its replicating portfolio by testing for unit roots, by means of Augmented Dickey-Fuller tests (ADF). Second, we check whether there is a common long-term trend in the swap and in the replicating portfolio by using cointegration analysis, and in particular, we apply the Johansen cointegration test. Third, we analyse price convergence between the swap price and the price of its replicating portfolio to determine to what extent persistent differences exist. If the price of the replicating portfolio differs systematically from the price of the swap contract, this can be due to some impediment or transaction costs that prevent full intertemporal integration. We test for convergence by estimating the models (Borenstein, et al., 2008).

$$Q_{bc}(t) - AQ_{bc}(t) = \alpha + \varepsilon(t) \quad , b=1,2,\ldots,B \text{ and } c=1,2,\ldots,C \quad (3)$$

$$Y_f(t) - AY_f(t) = \alpha + \varepsilon(t) , f=1,2,\ldots,F \quad (4)$$

If replicating portfolios $AQ_{bc}$ and $AY_f$ are unbiased estimations of swap prices $Q_{bc}$ and $Y_f$ respectively, then we should expect that $\alpha = 0$. Fourth, we also perform the variance ratio test between (the returns of) swaps and replicating portfolios. This test measures the size of basis risk, that is, to what extent the difference between fundamental price (e.g. Q11) and the price of the hedging portfolio (e.g. AQ11) is volatile. Alternatively, whether the volatility of innovations of fundamental price is equal to the volatility of innovations of replicating portfolio. The lower the size of this basis risk, the higher the time-zero efficiency.

## 2.3 Trading rules

To gain further insight on the degree of time-zero efficiency, we study whether a trader would have been able to capitalize on persistent price differences. Following Borenstein et al. (2008), we consider simple trading rules and we evaluate whether they would have made profits during the full period, as well as in some sub periods. Institutional impediments and traders' incomplete understanding of markets could create profitable trading rules. Additionally, we examine frequency of violations of simple no-arbitrage restrictions between basic contracts and their replicating portfolios, both excluding and including transaction costs

The first trading rule we consider assumes that a trader uses recent price differences between the swap and its replicating portfolio. We assume that during every day t a trader purchases the contract that was less expensive the previous day (t-1) and sells an equal size of the more expensive contract (as observed at t-1). We assess whether this rule would have produced profits in the hands of a pure arbitrageur, unconstrained by limits in his arbitrage activities. In the second trading rule, we assume that during every day *t*, if there is a difference in price between a swap contract and its replicating portfolio, the trader purchases the contract that is less expensive and sells an equal size of the more expensive contract. In other words, we examine the frequency of zero-time no-arbitrage violations, both excluding and including transaction costs.

3. Data

Germany is Eurozone's largest power market. The data set consists of daily data from June 1, 2004 until December 31, 2012, on settlement prices for the following baseload[i] swap contracts traded in EEX: Yearly, Quarterly and Monthly (Phelix-Base/Month/Quarter/Year-Futures). The company operating EEX market (EEX AG) has provided the data.[j] We choose six liquid contracts within each market segment, that usually are the closest to maturity ones. Within each market segment, these six contracts represent 99% (100%), 97% (99%) and 100% (100%) of total trading volume (open interest, defined as the total number outstanding of derivative contracts that have not been settled) in the case of monthly, quarterly and yearly contracts, respectively. France is Europe's second largest power market. Our data set consists of daily data from July 3, 2006 to March 19, 2014 on settlement prices for the same contracts as in Germany. We choose six liquid contracts, these six contracts represent 99% (100%), 97% (99%) and 100% (100%) of total trading volume (open interest) in the case of monthly, quarterly and yearly contracts, respectively. In Germany and France monthly, quarterly and yearly contracts account, on average, for 99% of total volume of futures contracts, whereas shorter-term contracts account for around 1% of total volume (EEX, 2013).

---

[i] Specifically, we use EEX Power Derivatives Phelix-Base Month, Quarter and Year Future (Exchange codes F1BM, F1BQ, F1BY). As an illustration of this kind of contracts, the 1MW baseload Jan13 contract is a monthly swap contract that gives the holder the obligation to buy 1MWh of energy for each hour of January 2013, paying the futures price in Euros/MWh. The seller provides the buyer the amount of energy of 1MW × 24h × 31. The settlement is financial.

[j] The futures market at the EEX started trading financial futures on base and peak block contracts in the spring of 2001. In 2004 options trading on these contracts were introduced, and since 2005, futures with physical settlement were introduced. European Energy Exchange (2005) and Viehmann (2011) describe the basic facts of the EEX futures market.

Spain is Europe's fifth largest power market. Our data set consists of daily data from July 3, 2006 until June 6, 2014 on settlement prices for the following baseload[k] swap contracts traded in OMIP: Yearly, Quarterly and Monthly. The Iberian Forward Market Operator (OMIP) has provided the data.[l] We choose liquid contracts within each market segment, that usually are the closest to maturity ones. Within each market segment, we select five monthly contracts, six quarterly contracts and three yearly contracts, these contracts representing 100% (100%), 100% (100%) and 100% (100%) of total trading volume (open interest) in the case of monthly, quarterly and yearly contracts, respectively. Monthly, quarterly and yearly contracts account, on average during the sample, for around 95% of total volume of futures in OMIP market, whereas shorter-term (less than one month) contracts account for around 5% of total volume[m]. Liquidity in this market is considerably lower than in other markets, for instance trading volume is on average 5% of German market and slightly lower than French market.

We define continuous series as a perpetually linked series of swap settlement prices. For example, M1 starts at the nearest contract month, which forms the first values for continuous series, until contract reaches its expiry date, or until the first business day of the actual contract month. At this point, we take the next trading contract month.[n]

---

[k] Specifically, we use OMIP's FTB (M,Q,Y) which are Base load futures for Spain, with daily settlement (Exchange codes FTBM, FTBQ, FTBYR). The characteristic of the contract is similar to the one in footnote h. Settlement is financial.

[l] The OMIP market started trading financial futures on base and peak block contracts on July 3, 2006. The evolution of trading in this market is explained in Capitán Herráiz, and Rodríguez Monroy (2012), Furió and Meneu (2010) and Capitán Herráiz (2014).

[m] During the last years, there has been an increase in the volume of shorter-term contracts. According to recent CNMC bulletins (www.cnmc.es/es-es/energia/mercadosderivados) in 2103 (2014) shorter-term (less than one month) contracts accounted for 8.9% (10.4%) of total volume. However, during initial years of the sample, volume of short-term contracts was around 1-2%. For instance, daily and weekend contracts were not available in OMIP until May 2011.

[n] In the continuous series, a change in the level of the series appears on the day when one contract expires and a new one is included. This "rolling" effect sometimes generates jumps in the price series. To deal with this

Using notation described in section 2, we set maximum number of months as J=6, maximum number of quarters ahead in which contract delivers the electricity as B=2 and, given that c refers to the quarter in which the contract is traded, we set C=4. Finally, we set F=1 as maximum number of years ahead on which we build replicating portfolios. We show definitions for all replicating portfolios we employ in this paper in Table 1. Figure 1 presents graphically an example of interaction between different swap contracts and their replicating portfolios.

[INSERT TABLE 1 HERE]

[INSERT FIGURE 1 HERE]

As may be seen in Figure 1, in January, buyer of contract Q11 obtains a supply of electricity at fixed price during April, May and June, exactly in the same terms as the buyer of contracts M3, M4 and M5. Therefore, replicating portfolio of Q11 contains these three contracts (in January). In February, buyer of contract Q11 obtains a supply of electricity at fixed price during April, May and June, exactly in the same terms as the buyer of contracts M2, M3 and M4. Therefore, specific monthly contracts needed to set up the replicating portfolio for Q11, change over time.

*3.1 Summary Statistics*

---

artificial effect, we apply intervention analysis in each day when there is a "rolling" effect. Notice that market behavior does not cause these jumps. They are simply a technical problem caused by the definition of the continuous series.

Table 2 provides information on basic statistics for all series included in efficiency tests, using the full sample. Swap data are in Panel A and replicating portfolios are in Panel B. Average values of the swap contracts are lower than average values of replicating portfolios in 18 out of 23 cases (78%) and volatilities of swap contracts are usually lower than volatilities of replicating portfolios. It is also worth noting that volatility decreases with contract's maturity (this fact is known as the Samuelson effect), a phenomenon also documented by Capitán Herráiz (2014).Oher statistics related with the shape of the distribution (skewness, kurtosis) are similar, suggesting a relatively homogeneous behaviour in all cases[o].

[INSERT TABLE 2 HERE]

Table 3 provides the same information as in Table 2, but the sample includes only days with positive trading volume. In the case of replicating portfolios, this implies that all components of the portfolio must present a positive trading volume. Some differences appear in comparison with Table 2. Average values of swap contracts are lower than average values of replicating portfolios in 10 out of 17 cases (59%). Notice that there are several contracts in some markets whose replicating portfolios can be constructed only in few days, for instance Q12 in France (6 days) and Spain (9 days). However in the case of yearly contracts in all markets, as well some quarterly contracts in the German market (e.g. Q12, Q13 and Q14), there is enough sample size to allow for a meaningful analysis.

---

[o] There are some larger differences in the case of the French market. In the first subsample, from 2006 to April 2009, there are six monthly contracts available, whereas in the second subsample, from April 2009 to the end of the sample, there are only three monthly contracts available (M1, M2 and M3). This fact obviously affects the construction of replicating portfolios. We computed all statistics using the first subsample and statistics are much more homogeneous and similar to Germany. Detailed results are available on request.

[INSERT TABLE 3 HERE]

This situation appears because trading volume tends to concentrate in a subset of contracts as can be seen in Figure 2 corresponding to Germany, France and Spain respectively. In the German market, Y1 contract is the liquid contract by far, because it accounts for 82% of total volume in period 1, and for 65% of total volume in all periods. Yearly contracts present higher trading volume in periods 1, 2 and 3. Quarterly contracts are relatively liquid in periods 4, 5, and 6. Monthly contracts present sizeable trading volume only for periods 1, 2 and 3. In the French market, the liquid contract is Y1 which accounts for almost 70% of trading volume, followed by Q1 (21%) and by M1 (9%). The situation is different in the Spanish market because Q1 contract is the liquid contract by far[p], because it accounts for 57% of total volume in period 1, and for 61% of total volume in all periods. The reasons for this peculiarity of OMIP market are explained in Capitán Herráiz (2014). From 2007 to 2013 a number of regulated forward contracting mechanisms were applied in this market, such as Virtual Power Plant (VPP) Auctions (EPE auctions), and Last Resort Supply Auctions (CESUR auctions). These auctions boosted the liquidity of quarterly contracts with the same maturity as those from EPE and CESUR auctions, possibly for risk management purposes. Notice that quarterly contracts present higher trading volume in all periods, but in periods 5 and 6 there is hardly any trading volume.

[INSERT FIGURE 2 HERE]

4. **Empirical Analysis**

---

[p] Notice that Figure 2 contains the average daily trading volume, including days where the trading volume is zero. Obviously, the values change if we compute the average using only days with non-zero volume. For instance in the case of Spain, the average volume for Y1 is 21213.63 MWh (all days, n=1436) and is 74869.43 MWh (days with non-zero volume, n=563).

**4.1 Unit root tests**

Table 4 shows unit root test results for swap contracts and replicating portfolios. Panel A reports results using all observations and Panel B the results taking into account only days with positive trading volume.

[INSERT TABLE 4 HERE]

We may observe from the table that, in almost all cases, we cannot reject the null hypothesis of unit root at conventional significance levels. From the efficiency point of view, the problematic situation appears when both the swap contract and the replicating portfolio are stationary. The higher the number of such cases, the more inefficient is the market. For Germany, there is only one case (Q14, AQ14) in which this situation happens in the full sample and in the volume-restricted sample there is none. In the cases of France and Spain also, in general, all contracts and replicating portfolios present unit roots but there are some cases in which the replicating portfolio does not have enough volume to allow for meaningful estimations. In summary, the results of unit root tests do not point out clear signs of inefficiency.

**4.2 Cointegration test**

Table 5 shows efficiency results as measured by cointegration tests. Panel A reports results for the full sample and Panel B the results taking into account only the days with positive trading volume[q]. From the efficiency point of view, the problematic situation appears when both the swap contract and the replicating portfolio are not cointegrated.

[INSERT TABLE 5 HERE]

From the table, we may observe that, in all markets and contracts, we cannot reject the null hypotheses of at least one cointegration relationship at conventional significance levels. The normalized cointegration coefficients are [1,-1], as expected. From the efficiency point of view, the problematic situation appears when we do not find cointegration between the swap contract and the replicating portfolio. As it was the case with unit root tests, the higher the number of such cases, the less efficient is the market. In the three markets, both in the full sample and in the volume-restricted sample, we cannot reject the null hypothesis of cointegration for each pair of contracts at conventional significance levels. In summary, the cointegration analysis suggests a similar conclusion as the one obtained from the results of unit root tests, in the sense that we do not find no clear signs of inefficiency.

**4.3 Convergence test**

Following the approach in Borenstein et al. (2008) to ascertain the extent of inefficiency between the forward market and the spot market, we analyse price convergence between the swap contract and its replicating portfolio. Persistent price differences between a given swap contract and its replicating portfolio would imply the existence of inefficiencies that could be exploited by arbitrageurs unrestricted by funding constraints (Shleifer and Vishny, 1997).

---

[q] Notice that replicating portfolios are discontinuous series. They can be computed only in some days of the sample as we may observe in Table 1. Missing values and the analysis of days with positive volume introduces restrictions across the different segments. Thus, in Table 5 we report the results of the cointegration analysis using the first segment of the common sample. Results for the other segments are similar and are available on request.

[INSERT TABLE 6 HERE]

We estimate equations (3) and (4) by means of OLS regressions based on Newey-West procedures and therefore standard errors of the regression coefficients are robust to heteroscedasticity and autocorrelation. We show results in Table 6. Panel A reports results for the full sample and Panel B results using days with positive trading volume. For the German case, in Panel A we may see that in five out of nine cases the prices of swap contracts are significantly lower than the prices of their corresponding replicating portfolios. These daily price differences range between 0.06 €/MWh (0.1%) and 0.008€/MWh (0.02%) on average. In panel B, we may observe that in six out of nine cases the prices of swap contracts are significantly lower than the prices of the corresponding replicating portfolios, the range of variation of these differences is similar to the one observed in panel A. However, in four cases, the regression contains less than 20 observations, which imply that the results should be taken with caution. On the other hand, in the other two cases (Q11 and Y1) the sample size is large enough to make the inference reasonably reliable. In four cases, results from panel A and Panel B coincide, and, in particular, they coincide for Q11 and Y1. Therefore, the evidence suggests that, at least in two cases, the price of the replicating portfolio was, on average, significantly higher than the price of the swap contract, suggesting lack of convergence.

In the French market and by using the full sample, we find that in one (two) case(s) prices of swap contracts are significantly higher (lower) than prices of their corresponding replicating portfolios. In panel B we observe that the only case in which the replicating portfolio can be computed, the Y1 swap contract, the price of this contract is significantly lower than the price of the corresponding replicating portfolio. Therefore, there is (at least) one case in which the swap price is lower than the price of the replicating portfolio. Finally, in the Spanish market, in Panel A, in one out of five cases, the prices of swap contracts are significantly lower than prices of their corresponding replicating portfolios. In panel B there is one case in which swap price is above the price of the replicating portfolio. Both findings suggest lack of convergence.

We should stress the fact that we are able to apply the tests of convergence with positive trading volume (containing at least 100 data points) in five cases out of nine (56%) in the German market, and in one case out of nine (11%) in the French and Spanish markets. This result suggests that the inclusion of realistic liquidity restrictions (e.g. requiring a minimum of trading days with positive trading volume, allowing for reasonable statistical inference) is a crucial factor when assessing the results from time-zero efficiency tests.

In any case, in all markets there are some significantly persistent deviations between prices of swap contracts and prices of their replicating portfolio. In most cases, the price of the replicating portfolio is above the price of the swap contract. The extent to which a trader can exploit this discrepancy to obtain economic profits is the subject of section 4.5.

**4.4 Variance ratio test**

Table 7 shows results of the variance ratio tests, as measured by the F statistic. Panel A reports results for the full sample and Panel B results taking into account only the days with positive trading volume in all markets. From the efficiency point of view, the problematic situation appears when the volatility of the returns of swap contract and the volatility of the returns of the replicating portfolio are different, because the size of this difference is directly related with the size of basis risk[r]. The higher the size of this basis risk, the lower the time-zero efficiency.

[INSERT TABLE 7 HERE]

In Germany and Spain, the results using the full sample essentially suggest that the variance ratio is very close to one, which implies a very low basis risk. In the case of France, there are many variance ratios different from one. This result is related with the problem pointed out in footnote *l* because in the first subsample, from 2006 to April 2009, there are six monthly contracts available, whereas in the second subsample, from April 2009 to the end of the sample, there are only three monthly contracts available (M1, M2 and M3). This fact obviously affects the construction of replicating portfolios. We computed all statistics using the first subsample and statistics are much more homogeneous and similar to Germany and Spain. Detailed results are available on request. On the other hand, when using the volume-restricted sample, in two cases (Germany), the volatility of the swap contract is substantially lower than the volatility of the replicating portfolio, suggesting some basis risk, and in one case (Spain), the volatility of the swap contract is substantially higher than the volatility of the replicating portfolio, suggesting some basis risk. Therefore, as it was the case with the convergence test, the evidence points to some possible instances of inefficiency.

---

[r] Usually, the basis is defined as the difference between the hedging instrument (e.g. futures) and basic (cash) price. In our setting however, we consider the swap contract as the basic price and the replicating portfolio as the hedging instrument. Then, the basis is the difference between the price of the basic swap contract (e.g. Y1) and its replicating portfolio (e.g. AY1).

## 4.5 Trading rules

Although results of convergence tests point out several cases of persistent price misalignments, and results from variance ratio tests suggest the existence of significant basis risk in a number of cases, it is unclear whether a trader would have been able to capitalize on the significant price differences uncovered in previous sections. To gain insight on this point, we consider some trading rules and evaluate whether they are profitable during the full period we analyse. To make them more realistic, we present results excluding or including trading costs.

In the first trading rule (Price Differences, PD henceforth), a trader buys the contract that was less expensive the previous day and sells an equal size of the more expensive contract (previous day). In the second trading rule (Arbitrage Violations, AV henceforth), if there is a difference in price between the swap contract and its replicating portfolio (a violation of the no-arbitrage condition), during every day $t$ a trader purchases the contract that is less expensive and sells an equal size of the more expensive contract. In doing so, we are able to examine the frequency of no-arbitrage violations.

[INSERT TABLE 8 HERE]

Table 8 summarizes average daily profits, t-statistics and cumulative profits for PD trading rule without transaction costs. The price difference strategy without transaction costs produces positive and statistically significant daily profits for seven out of nine swap contracts (Germany), for four out of nine contracts (France) and for five out of five contracts (Spain). Therefore, a trader that follows this strategy obtains significant economic profits in all markets. This strategy boils down to buying the swap contract and selling the replicating portfolio, in most cases. As an illustration, in the case of Germany in 86% (14%) of days, the trader buys (sells) swap contract and sells (buys) replicating portfolio. Failures in price convergence between swap prices and replicating portfolios reported in section 4.3 and Table 8 may appear if there are differential costs associated with contracting the assets. Trading and clearing fees amount to a flat rate of 0.01€/MWh for each transaction on average across markets. Average bid-ask spreads[s] for Germany are 0.8% (Monthly), 0.6% (Quarterly) y 0.3% (Yearly) (EEX, 2012, 2013,2014), for France are 2.7% (Monthly), 1.9% (Quarterly) and 1.4% (Yearly). In the OMIP, they are 3.2% (Monthly), 2.6% (Quarterly) and 1.9% (Yearly) (Capitán Herráiz and Rodríguez Monroy, 2009). Table 9 presents the same analysis considering transaction costs. When realistic transaction costs are included, PD rule produces statistically and economically significant losses over the sample period in all markets.

A derivatives market is time-zero efficient when all the relevant and ascertainable information is fully and immediately reflected in the market prices of the derivatives contracts. Therefore, there are no time-zero arbitrage opportunities. An arbitrage opportunity exists if it is possible to design a costless strategy that can yield riskless economic profits, using only contracts available at time-zero. Traditionally, we evaluate market efficiency by testing the existence of predictable patterns in the prices. For instance, if there are two contracts A and B that provide essentially equivalent cash flows, if there are no frictions, their prices should be equal in every period. However, if prices follow predictable patterns (e.g., the price of A tends to be lower than the price of B), agents could use this information to design trading strategies that yield abnormal profits. Nevertheless, existence of frictions (transactions costs) changes the situation. A trading strategy that seems to provide riskless profits may cease to be profitable once transactions costs are taking into account. In the case of the German market, we document this situation to a considerable extent. In conclusion, although we find some predictable patterns, they cannot provide profitable arbitrage opportunities and we conclude that this evidence is consistent with the time-zero efficiency of the German market. Evidence is less compelling in the case of the French and Spanish markets, however.

---

[s] The average bid-ask spread is computed as *(Ask -Bid)/[(Ask+Bid)/2]*. We report results based on average bid-ask spreads for basic contracts and for contracts in the replicating portfolio. Detailed results are available on request.

[INSERT TABLE 9 HERE]

Table 9 summarizes results from AV trading rule. We present results for the full sample in Panel A, and, in Panel B results for the restricted sample, including only days with positive trading volume. Column "No Trans. Cost" contains the percentage of violations of no-arbitrage conditions, assuming no transaction costs. Column "Sell Repl. Portf." contains the percentage of violations observed by selling the replicating portfolio and buying the basic contract, and including transaction costs. Column "Buy Repl. Portf." contains the percentage of violations observed by buying the replicating portfolio and selling the basic contract, and including transaction costs. In the full sample, the AV test without transaction costs indicates persistent violations of no-arbitrage conditions in all markets and all contracts, ranging from 50% to 100%. In the restricted sample (only days with positive trading volume), the percentage of violations range from 26% to 100%. However, when we take into account transaction costs, in the full sample, the violations range between 0% and 1.85%. In the restricted sample, the violations range from 0% to 1.26%. In summary, the empirical evidence suggests a very low frequency of no-arbitrage violations beyond transaction costs. This fact is consistent with zero-time efficiency of all markets.

## 5. Liquidity and Representativeness

In this section, we present some characteristics of liquidity and representativeness of the markets, and we discuss the extent to which these characteristics may affect our results. Regarding liquidity[t], in the three markets trading volume tends to concentrate in a subset of contracts (usually, near term contracts), as we shown in Figure 2. This situation is not uncommon in other derivatives markets. In addition, in the German market liquidity tends to be evenly spread across the twelve months of the year. However, this is not the case in the French and Spanish markets, where there are remarkable differences between months. In Figure 3 we present the differences (in %) of each month with respect to the month with lowest number of trading days. We define trading days as those where there is positive trading volume in at least one contract. In the case of the German market, December is the month with fewer trading days (56% of the days there is a trade in at least one contract) and February is the month with more trading days (65%). In the cases of the French and Spanish markets, the months with fewer trading days are February (6%) and September (8%) respectively, and the months with more trading days are June (12%) and February (17%) respectively. In Figure 4 we present average differences (%) of all months with respect to the month with lowest number of trading days. In the case of the German market, the difference is 7.88% that suggest that the number of trading days tends to be relatively similar across months. However, this is not the case in the French market, where the difference is 38.83%. The extreme case is the Spanish market, in which the difference is 49%, suggesting large differences in trading activity between months. We discuss the policy implications of these facts in the next section.

---

[t] The sources of this section are as follows. On the German Market, we use monthly reports by EEX on market activity (www.eex.com) from June 2004 to December 2014 and the BUNDESNETZAGENTUR (www.bundesnetzagentur.de) Monitoringbericht from 2004 to 2014. In the case of the Spanish market, we use the reports by Comision Nacional de los Mercados y la Competencia (www.CNMC.es) "Informes de seguimiento (supervisión) de mercados a plazo de energía eléctrica en España", from June 2007 to December 2014. From June 2007 to May 2013 the reports were elaborated by the Comision Nacional de la Energia (CNE), later subsumed into the CNMC, Comision Nacional de los Mercados y la Competencia, on October 2013. In the case of the French market we use reports published by the Commission de Regulation de L'Energie (www.CRE.fr), specifically the "Electricity and gas market observatory reports" from 4th quarter 2004 to 4th quarter 2014.

Regarding representativeness, activity in electricity derivatives markets takes place in two segments. Organized Market Places (OMP) using standard contracts manages the first segment. The second segment takes place over-the-counter (OTC), through direct transactions or through intermediaries (brokers and trading platforms). In the case of OMP, transaction prices and trading volume are transparent and publicly available. In addition, regulators have access to detailed data. The European Union Regulation on Market Integrity and Transparency of Wholesale Energy Markets (REMIT)[u], empowers those authorities to access all transactions data. On the other hand, information of OTC trading and volume is provided to regulators, on a voluntary basis, by some brokers and agents in the OTC market. They send (weekly or monthly) reports to regulators in which they describe contracts traded and closing prices (possibly with some information on bid-ask spreads), but they do not report information on the identity of the counterparties. The reliability of tests on zero-time efficiency depends crucially on the representativeness of prices used. Ideally, we should use the prices of the market segment (OMP or OTC) which has highest liquidity. However, only OMP prices are publicly available. Thus, we use these in this paper. In Figure 5 we present the % of total volume in electricity derivatives markets traded in OMP for the German, French and Spanish markets. In the case of the German market, the average market share of OMP is 37% and there is a steady increasing trend, suggesting that OMP prices are likely to be representative of the overall situation of the German electricity derivatives market. However, in the French market, average market share is 4.6% and it does not follow a clear trend, and in the Spanish market, average market share is 15%, but the trend was decreasing between 2007 and 2010, it increases between 2010 and 2013, and presents a slight decrease in 2014. These facts suggest that, in the cases of the French and Spanish market, OMP prices are less likely to be representative of the overall situation of their respective electricity derivatives markets than in the case of the German market. We discuss the policy implications of these findings in the next section.

---

[u] Applied since the 28th of December 2011

## 6. Conclusions and Policy Implications

We present a method to assess the degree of time-zero efficiency of electricity derivatives markets using derivatives contracts only. Using this method, this paper presents new results on the evaluation of time-zero efficiency of derivatives power markets in Germany, France and Spain. We derive several conclusions from our results. First, a battery of different tests and trading rules may be needed to properly assess time-zero efficiency in a given derivatives market. The reason is that different tests may produce conflicting results. Second, inclusion of realistic liquidity restrictions (e.g. positive trading volume) is a crucial factor when assessing results from time-zero efficiency tests. Importantly, we are able to apply tests of convergence with positive trading volume (containing at least 100 data points) in the German market in five cases out of nine (56%), and in the French and Spanish markets in one case out of nine (11%). Third, transaction costs must be taken into account when assessing the performance of alternative trading rules. In fact, simple trading strategies that appear to be highly profitable in gross terms produce significant losses when we take into account liquidity availability and bid-ask spreads.

A first policy implication is that the bulk of the evidence suggest a substantial degree of time-zero efficiency in the German power derivatives markets, which is good news, given policy concerns about market structure and efficiency (EC, 2007). However, as we report in section 5, there are significant differences in liquidity and representativeness across markets.

On the liquidity side, in the German market the number of trading days tends to be relatively similar across months. On the other hand, in the French and Spanish markets we document large differences in trading activity between months. This concentration of liquidity in some months and trading sessions has the potential of market abuse from excessive speculation in the spot and/or futures market. Thus, a second policy implication (see also the recommendations in Chapter 8 of Capitán Herráiz (2014)) is the need to increase liquidity in the French and Spanish power derivatives markets, by means of procedures that give incentives to traders to spread liquidity evenly over the twelve months. One method is through marketing incentives (e.g. auctions in some months, fee discounts) to attract financial companies, hedge funds, commercial banks, government-sponsored enterprises, energy-intensive industries, renewable generation companies and other players (many of them already active in the spot market) attracted by trading opportunities.

On the representativeness side, in the German market, average market share of OMP is 37% and increasing. However, in the French and Spanish markets, average market share is 4.6% and 15% respectively, and steady or decreasing. This low representativeness of OMP in France and in Spain should concern regulators. First, overall market transparency may be compromised. Second, lack of post-trade transparency and centralised clearing does not facilitate wholesale market supervision. Third, retail price formation in liberalised markets may be based on wholesale price signals that may potentially differ.[v] Thus, a third policy recommendation is to encourage OMP administrators to attract OTC trading towards OMP venues, by using pricing and marketing incentives. In addition, clearing houses should facilitate the widest possible array of maturities for futures trading, in order to attract players with different time horizons. Furthermore, clearing houses should provide incentives to settle OTC-traded volumes in an organized clearinghouse

REMIT regulations empower European regulators to access all wholesale trading data (transactions and orders to trade) in OMP, as they are to be reported by market participants to the Agency for the Cooperation of Energy Regulators (ACER). However, there is not a similar requirement in the case of trading out of OMP (e.g. OTC without using broker platforms, as pure bilateral trades), as orders are not to be reported to ACER. Our fourth policy recommendation is therefore to encourage the collection and publication of statistics of prices and trading volumes also in OTC markets[w]. This information would increase ex-post trade transparency and efficiency of European wholesale energy markets. In this vein, it should be noted that neither European OMPs nor European clearing houses for energy derivatives publish statistics regarding long and short positions associated to commercial traders (i.e. energy companies) or to financial entities. In this regard, an interesting initiative to be taken into account is the information on long, short and spread positions by types of traders (e.g. "commercial" and "non-commercial" ones) found in the Commitment of

---

[v] We have not found any published study of zero-time efficiency between contracts negotiated in OMP and in OTC in European power derivatives markets.
[w] The recent Commission Implementing Regulation (EU) No 1348/2014 of 17 December 2014, regarding data reporting on wholesale energy market, sets out what are the requisites of information regarding wholesale trades in electricity and gas to be provided by agents to ACER and to the national regulatory authorities. However, as far as we know, this regulation applies to OMP only.

Traders (COT) reports, published by the U. S. Commodity Futures Trading Commission (CFTC). Publication of such statistics by each clearing house, or by regulatory agencies having access to centrally cleared data , would provide more post-trade transparency, increasing confidence in the market and thus its efficiency.

Finally, whilst the use of several techniques in this paper has provided useful insights, there are some limitations related to the kind of contracts employed in the analysis. Shorter-term contracts (e.g. weekly) and contracts based on peak prices should also be taken into account in order to obtain a fuller perspective on the issue of time-zero efficiency. Looking forward, application of our methodology to other power derivatives market is an immediate extension, which we left for future research.

# References

Benth, F. E., A. Cartea, and R. Kiesel (2008). Pricing forward contracts in power markets by the certainty equivalence principle: Explaining the sign of the market risk premium. Journal of Banking and Finance 32, 2006-2021.

Bierbrauer,M., M. Menn, S.T. Rachev and S. Trück (2007). Spot and derivative pricing in the EEX power market. Journal of Banking and Finance 31, 3464-348.

Boisselau, F., (2004). The role of power exchanges for the creation of a single European Electricity market: market design and market regulation. Delft University Press.

Borenstein, S., J. Bushnell, C.R. Knittel and C. Wolfram (2008). Inefficiencies and market power in financial arbitrage: A study of California's electricity markets. Journal of Industrial Economics, 56, 2, 347-378.

Bower J., (2002). Seeking the Single European Electricity Market: Evidence from an Empirical Analysis of Wholesale Market Prices, Working Paper. Oxford Institute for Energy Studies.

Benth, F.E., and S. Koekebakker (2008). Stochastic Modeling of Financial Electricity Contracts. Energy Economics, 30, 1116–1157.

Bunn, D.W. and A. Gianfreda (2010). Integration and shocks transmission across European electricity forward markets. Energy Economics, 32, 278-291.

Capitán Herráiz, A. (2014). Regulatory proposals for the development of an efficient Iberian energy forward market. Ph.D. Dissertation. ETSII. Universidad Politécnica de Madrid.

Capitán Herráiz, A. and C. Rodríguez Monroy (2009). Evaluation of the Liquidity in the Iberian Power Futures Market. IV Congress of Spanish Association of Energy Economists (AEEE), Seville (Spain), January 2009.

Capitán Herráiz, A. and C. Rodríguez Monroy (2012). Evaluation of the trading development in the Iberian Energy Derivatives Market. Energy Policy, 51, 973-984.

Cartea, A. and P. Villaplana (2008). Spot price modeling and the valuation of electricity forward contracts: the role of demand and capacity. Journal of Banking and Finance, 32, 2502-2519.

European Commission (2007). DG Competition report on energy sector inquiry (SEC(2006) 1724, 10 January 2007).

European Commission (2009). Ensuring efficient, safe and sound derivatives markets. (COM(2009) 332).

European Commission (2014). Quarterly Report on European Electricity Markets. Vol 7, Issue 3.

European Energy Exchange (2005). EEX-Terminmarktkonzept, European Energy Exchange (EEX).

European Energy Exchange (2012). EEX Summer Workshop 2012, European Energy Exchange (EEX).

European Energy Exchange (2013). EEX Summer Workshop 2013, European Energy Exchange (EEX).

European Energy Exchange (2014a). EEX Summer Workshop 2014, European Energy Exchange (EEX).

European Energy Exchange (2014b). EEX- Kontraktspezifikation, European Energy Exchange (EEX).

Furió, D. and V. Meneu (2010). Expectations and forward risk premium in the Spanish deregulated power market. Energy Policy, 38, 784-793.

Lucia, J., and E.S. Schwartz (2002). Electricity prices and power derivatives: evidence from the Nordic Power Exchange. Review of Derivatives Research 5, 5–50.

Shahidehpour, M., H. Yamin and Z. Li (2002). Market Operations in Electric Power Systems: Forecasting, Scheduling, and Risk Management. Wiley.

Shleifer A., and R. Vishny (1997). The limits of arbitrage. Journal of Finance, 52, 35–55

Viehmann, J. (2011). Risk premiums in the German day-ahead Electricity market. Energy Policy, 39, 386-394.

Wilkens, S., and J. Wimschulte (2007). The pricing of electricity futures: evidence from the European energy exchange. Journal of Futures Markets, 27 (4), 387–410.

Zachmann G., (2008). Electricity wholesale market prices in Europe: Convergence? Energy Economics, 30 (4), 1659–1671.

# Table 1: Replicating Portfolios

This table reports replicating portfolios for three market segments (M,Q,Y) and eighteen contracts considered (M1,…,M6; Q1,…,Q6; Y1,…,Y6). Monthly contracts are M$j(t)$ where j=1,2,…,6. , quarterly contracts are Q$bc(t)$ where b=1,2 and c=1,2,3,4 and yearly contracts are Y$f(t)$ where f=1.

| January | February | March | April | May | June |
|---|---|---|---|---|---|
| AQ11=1/3(M3+M4+M5) | AQ11=1/3(M2+M3+M4) | AQ11=1/3(M1+M2+M3) | AQ12=1/3(M3+M4+M5) | AQ12=1/3(M2+M3+M4) | AQ12=1/3(M1+M2+M3) |
| | | AQ21=1/3(M4+M5+M6) | | | AQ22=1/3(M4+M5+M6) |
| | | | AY1=1/4(Q32+Q42+Q52+Q62) | AY1=1/4(Q32+Q42+Q52+Q62) | AY1=1/4(Q32+Q42+Q52+Q62) |
| **July** | **August** | **September** | **October** | **November** | **December** |
| AQ13=1/3(M3+M4+M5) | AQ13=1/3(M2+M3+M4) | AQ13=1/3(M1+M2+M3) | AQ14=1/3(M3+M4+M5) | AQ14=1/3(M2+M3+M4) | AQ14=1/3(M1+M2+M3) |
| | | AQ23=1/3(M4+M5+M6) | | | AQ24=1/3(M4+M5+M6) |
| AY1=1/4(Q23+Q33+Q43+Q53) | AY1=1/4(Q23+Q33+Q43+Q53) | AY1=1/4(Q23+Q33+Q43+Q53) | AY1=1/4(Q14+Q24+Q34+Q44) | AY1=1/4(Q14+Q24+Q34+Q44) | AY1=1/4(Q14+Q24+Q34+Q44) |

## Table 2: Summary Statistics

This table reports some descriptive summary statistics for daily swap prices. Swap data is in Panel A and replicating portfolios are in Panel B. For the German Market the full sample period goes from 6/1/2004 to 12/31/2012. For the French Market is from 6/1/2006 to 3/19/2014 and for the Spanish market from 7/3/2006 to 6/6/2014. The sample size is N in each case. N for the replicating portfolios depends on available data of monthly and quarterly swaps series. For the French Market M4, M5 and M6 are not available since 2009.

| | German Market (EEX) | | | | | | | | | French Market (EEX) | | | | | | | | | Spanish Market (OMIP) | | | | |
|---|---|---|---|---|---|---|---|---|---|---|---|---|---|---|---|---|---|---|---|---|---|---|---|
| **Panel A: Swap contracts** | | | | | | | | | | | | | | | | | | | | | | | |
| | Q11 | Q12 | Q13 | Q14 | Q21 | Q22 | Q23 | Q24 | Y1 | Q11 | Q12 | Q13 | Q14 | Q21 | Q22 | Q23 | Q24 | Y1 | Q11 | Q12 | Q13 | Q14 | Y1 |
| **Mean** | 43.55 | 48.07 | 54.20 | 54.80 | 45.73 | 53.97 | 57.64 | 44.76 | 52.29 | 41.29 | 46.35 | 61.97 | 64.02 | 43.86 | 62.06 | 65.50 | 44.36 | 55.47 | 44.84 | 51.30 | 51.27 | 52.06 | 51.23 |
| **Median** | 43.76 | 45.23 | 50.05 | 52.86 | 42.58 | 49.51 | 52.26 | 45.63 | 51.73 | 39.28 | 41.63 | 57.82 | 59.17 | 39.16 | 54.75 | 60.16 | 45.79 | 53.30 | 45.00 | 52.00 | 48.35 | 52.73 | 51.00 |
| **Maximum** | 61.24 | 86.09 | 97.50 | 94.03 | 64.45 | 92.90 | 95.00 | 55.22 | 90.15 | 61.25 | 105.32 | 114.47 | 119.13 | 74.19 | 109.50 | 115.41 | 53.25 | 93.29 | 63.90 | 66.46 | 75.08 | 72.50 | 75.10 |
| **Minimum** | 28.69 | 29.50 | 34.43 | 34.04 | 31.99 | 34.57 | 37.50 | 31.62 | 33.12 | 28.36 | 29.56 | 42.63 | 39.00 | 32.92 | 47.73 | 34.22 | 33.73 | 41.53 | 29.13 | 36.20 | 37.66 | 36.60 | 38.40 |
| **Std. Dev.** | 9.16 | 10.87 | 14.65 | 11.70 | 10.03 | 13.49 | 14.70 | 6.35 | 10.43 | 8.01 | 12.47 | 16.30 | 13.27 | 9.96 | 15.15 | 18.44 | 5.45 | 10.43 | 7.94 | 7.37 | 9.69 | 7.38 | 6.57 |
| **Skewness** | 0.10 | 1.02 | 1.15 | 0.44 | 0.41 | 1.02 | 1.10 | -0.64 | 0.75 | 0.69 | 1.16 | 1.43 | 1.71 | 0.94 | 1.36 | 1.66 | -0.31 | 1.64 | 0.18 | 0.09 | 0.91 | -0.09 | 1.09 |
| **Kurtosis** | 1.74 | 3.70 | 3.80 | 3.39 | 1.75 | 3.80 | 3.76 | 2.80 | 4.45 | 2.62 | 4.40 | 4.39 | 6.39 | 2.74 | 3.98 | 5.09 | 2.24 | 5.80 | 2.15 | 2.43 | 3.04 | 2.97 | 4.95 |
| **N** | 492 | 498 | 573 | 546 | 160 | 177 | 177 | 162 | 1617 | 484 | 419 | 506 | 483 | 150 | 137 | 154 | 139 | 1405 | 485 | 463 | 506 | 471 | 1436 |
| **Panel B: Replicating portfolios** | | | | | | | | | | | | | | | | | | | | | | | |
| | AQ11 | AQ12 | AQ13 | AQ14 | AQ21 | AQ22 | AQ23 | AQ24 | AY1 | AQ11 | AQ12 | AQ13 | AQ14 | AQ21 | AQ22 | AQ23 | AQ24 | AY1 | AQ11 | AQ12 | AQ13 | AQ14 | AY1 |
| **Mean** | 43.59 | 48.06 | 54.21 | 54.86 | 45.75 | 53.98 | 57.70 | 44.75 | 52.30 | 42.85 | 49.94 | 65.91 | 69.27 | 43.15 | 71.30 | 81.69 | 47.93 | 56.50 | 44.78 | 51.36 | 51.27 | 51.94 | 51.25 |
| **Median** | 43.79 | 45.21 | 50.07 | 52.89 | 42.59 | 49.52 | 52.32 | 45.65 | 51.73 | 41.66 | 48.56 | 59.36 | 67.30 | 36.68 | 50.68 | 71.36 | 47.52 | 54.04 | 45.04 | 52.01 | 48.50 | 52.71 | 51.01 |
| **Maximum** | 61.22 | 86.11 | 97.55 | 94.13 | 64.45 | 92.95 | 95.14 | 55.25 | 90.19 | 61.54 | 87.79 | 119.90 | 119.27 | 62.25 | 109.86 | 115.58 | 53.29 | 93.35 | 64.22 | 66.47 | 75.09 | 72.56 | 75.11 |
| **Minimum** | 28.72 | 29.52 | 33.77 | 34.05 | 32.01 | 34.58 | 37.53 | 31.63 | 33.12 | 28.39 | 29.62 | 42.68 | 47.00 | 32.94 | 47.79 | 58.50 | 41.44 | 25.59 | 29.14 | 36.22 | 37.67 | 36.61 | 38.40 |
| **Std. Dev.** | 9.17 | 10.88 | 14.67 | 11.72 | 10.03 | 13.49 | 14.73 | 6.36 | 10.44 | 9.19 | 13.44 | 19.93 | 15.31 | 12.00 | 23.73 | 22.00 | 3.66 | 10.80 | 7.92 | 7.36 | 9.66 | 7.30 | 6.57 |
| **Skewness** | 0.10 | 1.02 | 1.15 | 0.45 | 0.40 | 1.02 | 1.10 | -0.63 | 0.75 | 0.26 | 0.44 | 0.88 | 1.11 | 0.81 | 0.16 | 0.57 | -0.23 | 1.45 | 0.18 | 0.09 | 0.92 | -0.13 | 1.09 |
| **Kurtosis** | 1.74 | 3.71 | 3.80 | 3.40 | 1.74 | 3.80 | 3.76 | 2.79 | 4.46 | 1.89 | 2.22 | 2.60 | 4.15 | 1.74 | 1.19 | 1.51 | 1.68 | 5.25 | 2.17 | 2.45 | 3.08 | 2.90 | 4.95 |
| **N** | 492 | 498 | 573 | 546 | 160 | 177 | 177 | 162 | 1617 | 322 | 254 | 295 | 269 | 56 | 39 | 57 | 68 | 1219 | 477 | 454 | 494 | 462 | 1428 |

## Table 3: Summary Statistics – Positive Trading Volume

This table reports some descriptive summary statistics for daily swap prices and replicating portfolios with positive trading volume. Swap data are in in Panel A and replicating portfolios are in Panel B. For the German Market the full sample period goes from 6/1/2004 to 12/31/2012. For the French Market is from 6/1/2006 to 3/19/2014 and for the Spanish market from 7/3/2006 to 6/6/2014. The sample size is N in each case. N for the replicating portfolios depends on the available data of volume series of the monthly and quarterly swaps. For the French Market M4, M5 and M6 prices and volume series are not available since 2009.

| | German Market (EEX) | | | | | | | | | French Market (EEX) | | | | | | | | | Spanish Market (OMIP) | | | | |
|---|---|---|---|---|---|---|---|---|---|---|---|---|---|---|---|---|---|---|---|---|---|---|---|
| **Panel A: Swaps contracts** | | | | | | | | | | | | | | | | | | | | | | | | |
| | Q11 | Q12 | Q13 | Q14 | Q21 | Q22 | Q23 | Q24 | Y1 | Q11 | Q12 | Q13 | Q14 | Q21 | Q22 | Q23 | Q24 | Y1 | Q11 | Q12 | Q13 | Q14 | Y1 |
| Mean | 43.65 | 48.14 | 54.43 | 55.07 | 46.20 | 55.65 | 59.45 | 45.64 | 52.31 | 43.85 | 50.43 | 61.65 | 58.35 | 53.37 | 64.52 | 61.89 | 44.76 | 53.87 | 44.51 | 51.50 | 49.97 | 51.51 | 49.61 |
| Median | 43.80 | 45.25 | 50.25 | 53.04 | 44.66 | 50.33 | 53.04 | 46.14 | 51.74 | 42.19 | 52.81 | 64.16 | 58.42 | 54.53 | 65.00 | 61.60 | 46.41 | 53.98 | 45.25 | 53.38 | 47.55 | 52.35 | 49.08 |
| Maximum | 61.24 | 86.09 | 97.50 | 94.03 | 64.45 | 92.90 | 95.00 | 55.22 | 90.15 | 61.25 | 66.31 | 68.44 | 70.08 | 69.25 | 70.25 | 67.57 | 47.29 | 61.65 | 63.38 | 65.35 | 74.25 | 72.50 | 74.50 |
| Minimum | 28.69 | 29.50 | 34.43 | 34.04 | 31.99 | 34.57 | 37.50 | 31.62 | 33.12 | 36.11 | 34.93 | 50.50 | 48.14 | 39.04 | 52.81 | 48.12 | 38.69 | 46.98 | 29.25 | 36.50 | 38.95 | 36.60 | 38.40 |
| Std. Dev. | 9.13 | 10.95 | 14.63 | 11.62 | 9.91 | 13.26 | 14.35 | 5.53 | 10.44 | 7.15 | 8.33 | 5.63 | 5.05 | 7.10 | 4.87 | 4.94 | 3.23 | 2.91 | 6.93 | 5.91 | 6.84 | 6.60 | 6.14 |
| Skewness | 0.09 | 1.00 | 1.15 | 0.45 | 0.38 | 1.07 | 1.19 | -0.59 | 0.75 | 0.74 | -0.40 | -0.62 | 0.22 | -0.23 | -1.20 | -0.82 | -1.02 | 0.27 | 0.12 | -0.32 | 1.67 | -0.09 | 1.05 |
| Kurtosis | 1.75 | 3.66 | 3.78 | 3.42 | 1.72 | 3.76 | 3.73 | 3.32 | 4.45 | 2.60 | 2.11 | 1.89 | 2.57 | 3.81 | 3.77 | 3.57 | 2.25 | 2.96 | 2.12 | 3.07 | 6.88 | 3.24 | 5.55 |
| N | 488 | 485 | 564 | 534 | 152 | 153 | 155 | 137 | 1608 | 53 | 89 | 92 | 78 | 17 | 33 | 22 | 11 | 325 | 222 | 202 | 181 | 191 | 420 |
| **Panel B: Replicating portfolios** | | | | | | | | | | | | | | | | | | | | | | | | |
| | AQ11 | AQ12 | AQ13 | AQ14 | AQ21 | AQ22 | AQ23 | AQ24 | AY1 | AQ11 | AQ12 | AQ13 | AQ14 | AQ21 | AQ22 | AQ23 | AQ24 | AY1 | AQ11 | AQ12 | AQ13 | AQ14 | AY1 |
| Mean | 44.128 | 50.024 | 54.651 | 55.154 | 52.445 | 58.545 | 53.933 | 45.918 | 53.512 | 38.65 | 55.04 | | | | | | | 53.73 | 43.401 | 51.861 | 47.847 | 51.712 | 49.594 |
| Median | 43.833 | 47.303 | 50.72 | 53.077 | 56.215 | 55.853 | 51.817 | 45.542 | 52.603 | 38.61 | 54.26 | | | | | | | 53.41 | 45.042 | 49.983 | 47.545 | 48.307 | 49.878 |
| Maximum | 61.22 | 86.113 | 91.95 | 94.13 | 61.45 | 87.99 | 67.357 | 48.28 | 90.185 | 39.36 | 63.52 | | | | | | | 61.15 | 52.497 | 65.583 | 50.867 | 64.367 | 69.258 |
| Minimum | 28.717 | 30.387 | 34.723 | 34.46 | 33.227 | 35.107 | 38.62 | 44.44 | 33.123 | 38.10 | 51.56 | | | | | | | 47.10 | 30.883 | 41.02 | 46.383 | 46.583 | 38.773 |
| Std. Dev. | 9.344 | 11.431 | 11.871 | 9.823 | 8.517 | 13.949 | 8.781 | 1.212 | 10.07 | 0.45 | 3.49 | | | | | | | 3.08 | 7.339 | 6.971 | 1.409 | 7.402 | 4.75 |
| Skewness | 0.182 | 1.066 | 1.304 | 0.94 | -1.172 | 1.036 | -0.05 | 0.817 | 0.879 | 0.39 | 1.75 | | | | | | | 0.09 | -0.319 | 0.497 | 1.032 | 1.211 | -0.367 |
| Kurtosis | 1.748 | 3.728 | 4.878 | 4.962 | 2.937 | 3.496 | 2.119 | 2.821 | 4.751 | 2.14 | 5.07 | | | | | | | 2.64 | 1.808 | 3 | 3.056 | 2.801 | 3.555 |
| N | 233 | 213 | 241 | 181 | 16 | 19 | 10 | 8 | 1247 | 6 | 9 | | | | | | | 178 | 10 | 9 | 10 | 5 | 247 |

## Table 4: Augmented Dickey- Fuller Test

This table reports the results of the augmented Dickey-Fuller test that a variable (on the left column) follows a unit-root process. The null hypothesis is that the variable contains a unit root, and the alternative is that a stationary process generated the variable. Columns report Dickey-Fuller test statistic, the MacKinnon approximate p-value, the number of lagged differences and the number of observations. The sample size is N in each case. The number of Lags is one.

| | German Market (EEX) | | | | | | | | | French Market (EEX) | | | | | | | | | Spanish Market (OMIP) | | | | |
|---|---|---|---|---|---|---|---|---|---|---|---|---|---|---|---|---|---|---|---|---|---|---|---|
| **Panel A: Full Sample** | | | | | | | | | | | | | | | | | | | | | | | | |
| | Q11 | Q12 | Q13 | Q14 | Q21 | Q22 | Q23 | Q24 | Y1 | Q11 | Q12 | Q13 | Q14 | Q21 | Q22 | Q23 | Q24 | Y1 | Q11 | Q12 | Q13 | Q14 | Y1 |
| t-stat | -0.083 | 1.266 | -0.87 | -3.242 | -0.925 | 2.09 | -1.007 | 0.023 | -1.855 | 0.825 | -0.384 | 0.774 | -3.518 | 0.883 | 0.099 | 1.497 | 0.182 | 0.336 | 1.599 | -1.531 | 0.204 | -1.466 | -1.774 |
| p-val. | 0.951 | 0.996 | 0.798 | 0.018 | 0.78 | 0.999 | 0.751 | 0.96 | 0.354 | 0.992 | 0.913 | 0.991 | 0.008 | 0.993 | 0.966 | 0.998 | 0.971 | 0.979 | 0.998 | 0.518 | 0.972 | 0.55 | 0.394 |
| N | 476 | 479 | 555 | 527 | 144 | 158 | 159 | 143 | 1562 | 276 | 223 | 288 | 272 | 76 | 72 | 78 | 74 | 789 | 466 | 444 | 486 | 454 | 1387 |
| | AQ11 | AQ12 | AQ13 | AQ14 | AQ21 | AQ22 | AQ23 | AQ24 | AY1 | AQ11 | AQ12 | AQ13 | AQ14 | AQ21 | AQ22 | AQ23 | AQ24 | AY1 | AQ11 | AQ12 | AQ13 | AQ14 | AY1 |
| t-stat | 0.06 | 1.26 | -0.911 | -3.256 | -0.923 | 2.089 | -1.022 | 0.068 | -1.855 | 0.319 | 1.762 | 0.857 | -2.954 | -0.504 | 1.963 | 1.564 | -1.334 | 0.621 | 1.465 | -1.534 | 0.254 | -1.385 | -1.777 |
| p-val. | 0.963 | 0.996 | 0.784 | 0.017 | 0.78 | 0.999 | 0.745 | 0.964 | 0.354 | 0.978 | 0.998 | 0.993 | 0.039 | 0.891 | 0.999 | 0.998 | 0.614 | 0.988 | 0.997 | 0.517 | 0.975 | 0.589 | 0.392 |
| N | 476 | 479 | 555 | 527 | 144 | 158 | 159 | 143 | 1562 | 176 | 134 | 161 | 147 | 28 | 20 | 30 | 36 | 684 | 448 | 426 | 462 | 431 | 1380 |
| **Panel B Trading Volume** | | | | | | | | | | | | | | | | | | | | | | | | |
| | Q11 | Q12 | Q13 | Q14 | Q21 | Q22 | Q23 | Q24 | Y1 | Q11 | Q12 | Q13 | Q14 | Q21 | Q22 | Q23 | Q24 | Y1 | Q11 | Q12 | Q13 | Q14 | Y1 |
| t-stat | -0.102 | 1.404 | -0.938 | -3.306 | -0.843 | 2.428 | -0.53 | -0.553 | -1.854 | -1.867 | -1.202 | 0.085 | 0.466 | 0.331 | -2.626 | | | -1.719 | 0.787 | -0.628 | 1.323 | 0.299 | -1.5674 |
| p-val. | 0.949 | 0.997 | 0.775 | 0.015 | 0.806 | 0.999 | 0.886 | 0.881 | 0.354 | 0.348 | 0.673 | 0.965 | 0.984 | 0.978 | 0.08 | | | 0.421 | 0.991 | 0.864 | 0.996 | 0.977 | 0.4999 |
| N | 466 | 446 | 531 | 503 | 128 | 120 | 122 | 105 | 1539 | 12 | 21 | 25 | 14 | 5 | 13 | | | 96 | 68 | 76 | 67 | 80 | 82 |
| | AQ11 | AQ12 | AQ13 | AQ14 | AQ21 | AQ22 | AQ23 | AQ24 | AY1 | AQ11 | AQ12 | AQ13 | AQ14 | AQ21 | AQ22 | AQ23 | AQ24 | AY1 | AQ11 | AQ12 | AQ13 | AQ14 | AY1 |
| t-stat | | | | | | | | | 0.054 | | | | | | | | | -0.511 | | | | | 0.0185 |
| p-val. | | | | | | | | | 0.963 | | | | | | | | | 0.889 | | | | | 0.96 |
| N | | | | | | | | | 996 | | | | | | | | | 56 | | | | | 77 |

# Table 5: Johansen test for co-integration

This table shows statistics used to determine the number of cointegration equations using Johansen's multiple-trace test method between the swap prices and the replicating portfolio (on the left column). The null hypothesis of the trace statistic is that there are no more than r cointegration relations. Panel A, presents the test using the first segment of observations given that our series are piecewise. Panel B uses only dates with positive trading volume. N indicates the number of included observations after adjustments. We present the critical values to compare the trace statistic at r = 0 (no cointegration) and r =1 (one cointegration relationship) in columns 3 and 5. If the Trace Statistic is higher than the critical value, we reject the corresponding null hypotheses at 95% confidence level. The last five columns present the normalized cointegrating coefficients (Coeff.) and their standard errors (s.e.). In all cases, we assume no deterministic trend in the data.

|  | Trace Statistic Rank 0 | 5% critical value | Trace Statistic Rank 1 | 5% critical value | N | 1st Coeff. | 2nd. Coeff. | 2nd Coeff. s.e. | Constant | Constant s.e. |
|---|---|---|---|---|---|---|---|---|---|---|
| **Panel A: Full Sample** | | | | | | | | | | |
| **German Market** | | | | | | | | | | |
| Q11 AQ11 | 23.665 | 20.262 | 6.570 | 9.165 | 48 | 1 | -1.003 | 0.002 | 0.117 | 0.059 |
| Q12 AQ12 | 24.042 | 20.262 | 5.260 | 9.165 | 59 | 1 | -1.000 | 0.000 | 0.029 | 0.016 |
| Q13 AQ13 | 22.774 | 20.262 | 7.301 | 9.165 | 62 | 1 | -1.002 | 0.002 | 0.101 | 0.083 |
| Q14 AQ14 | 24.799 | 20.262 | 7.982 | 9.165 | 58 | 1 | -0.996 | 0.003 | -0.120 | 0.122 |
| Q21 AQ21 | 25.109 | 20.262 | 3.447 | 9.165 | 15 | 1 | -1.001 | 0.000 | 0.050 | 0.012 |
| Q22 AQ22 | 21.344 | 20.262 | 3.745 | 9.165 | 18 | 1 | -0.993 | 0.001 | -0.215 | 0.031 |
| Q23 AQ23 | 27.463 | 20.262 | 5.603 | 9.165 | 17 | 1 | -1.010 | 0.004 | 0.393 | 0.160 |
| Q24 AQ24 | 32.283 | 20.262 | 3.667 | 9.165 | 16 | 1 | -0.956 | 0.005 | -1.394 | 0.161 |
| Y1 AY1 | 31.306 | 20.262 | 5.314 | 9.165 | 180 | 1 | -0.997 | 0.000 | -0.108 | 0.012 |
| **French Market** | | | | | | | | | | |
| Q11 AQ11 | 29.591 | 20.262 | 7.782 | 9.165 | 56 | 1 | -1.000 | 0.000 | 0.025 | 0.010 |
| Q12 AQ12 | 23.567 | 20.262 | 1.410 | 9.165 | 56 | 1 | -1.006 | 0.001 | 0.246 | 0.027 |
| Q13 AQ13 | 32.447 | 20.262 | 7.013 | 9.165 | 46 | 1 | -0.985 | 0.004 | -0.942 | 0.248 |
| Q14 AQ14 | 16.346 | 20.262 | 4.638 | 9.165 | 46 | 1 | -1.001 | 0.001 | 0.138 | 0.056 |
| Q21 AQ21 | 21.386 | 20.262 | 7.339 | 9.165 | 13 | 1 | -0.998 | 0.000 | -0.067 | 0.007 |
| Q22 AQ22 | 40.629 | 20.262 | 2.046 | 9.165 | 13 | 1 | -1.005 | 0.001 | 0.287 | 0.027 |
| Q23 AQ23 | 94.987 | 20.262 | 5.585 | 9.165 | 12 | 1 | -1.002 | 0.000 | 0.199 | 0.032 |
| Q24 AQ24 | 20.641 | 20.262 | 7.145 | 9.165 | 12 | 1 | -1.062 | 0.015 | 2.696 | 0.644 |
| Y1 AY1 | 22.867 | 20.262 | 1.771 | 9.165 | 118 | 1 | -0.998 | 0.001 | -0.083 | 0.054 |
| **Spanish Market** | | | | | | | | | | |
| Q11 AQ11 | 21.590 | 20.262 | 0.842 | 9.165 | 50 | 1 | -0.998 | 0.005 | -0.023 | 0.189 |
| Q12 AQ12 | 24.637 | 20.262 | 1.484 | 9.165 | 52 | 1 | -1.016 | 0.014 | 0.797 | 0.642 |
| Q13 AQ13 | 35.334 | 20.262 | 6.465 | 9.165 | 36 | 1 | -0.998 | 0.030 | -0.141 | 1.762 |
| Q14 AQ14 | 29.292 | 20.262 | 4.974 | 9.165 | 38 | 1 | -1.038 | 0.007 | 1.980 | 0.384 |
| Y1 AY1 | 27.549 | 20.262 | 1.007 | 9.165 | 111 | 1 | -1.020 | 0.016 | 1.080 | 0.855 |
| **Panel B: Positive Trading Volume** | | | | | | | | | | |
| **German Market** | | | | | | | | | | |
| Y1 AY1 | 23.252 | 20.262 | 4.461 | 9.165 | 33 | 1 | -0.993 | 0.001 | -0.313 | 0.033 |
| **French Market** | | | | | | | | | | |
| Y1 AY1 | 21.252 | 20.262 | 2.695 | 9.165 | 136 | 1 | -0.994 | 0.062 | -0.199 | 2.119 |
| **Spanish Market** | | | | | | | | | | |
| Y1 AY1 | 24.374 | 20.262 | 4.314 | 9.165 | 146 | 1 | -1.113 | 0.042 | 5.781 | 2.174 |

## Table 6: Convergence Test

This table report a test for convergence by estimating the models

$$Q_{bc}(t) - AQ_{bc}(t) = \alpha + \varepsilon(t), \quad b=1,2,\ldots,B \text{ and } c=1,2,\ldots,C$$
$$Y_f(t) - AY_f(t) = \alpha + \varepsilon(t), \quad f=1,2,\ldots,F$$

We report autocorrelation and heteroskedasticity-adjusted Newey-West standard errors (s.e). Panel A, presents tests using all observations. Panel B uses only dates with positive trading volume. Figures in boldface are significant at 1% confidence level.

| | German Market | | | | | French Market | | | | | Spanish Market | | | | |
|---|---|---|---|---|---|---|---|---|---|---|---|---|---|---|---|
| | α | s.e | t-stat | p-value | N | α | s.e | t-stat | p-value | N | α | s.e | t-stat | p-value | N |
| **Panel A: Full Sample** | | | | | | | | | | | | | | | |
| Q11 AQ11 | -0.039 | 0.010 | **-4.001** | 0 | 492 | -0.009 | 0.033 | -0.265 | 0.791 | 321 | 0.006 | 0.011 | 0.516 | 0.606 | 476 |
| Q12 AQ12 | 0.015 | 0.019 | 0.786 | 0.432 | 498 | 0.448 | 0.183 | **2.446** | 0.015 | 253 | -0.01 | 0.006 | -1.793 | 0.074 | 454 |
| Q13 AQ13 | -0.012 | 0.009 | -1.371 | 0.171 | 573 | -0.130 | 0.121 | -1.076 | 0.283 | 294 | -0.003 | 0.014 | -0.234 | 0.815 | 494 |
| Q14 AQ14 | -0.059 | 0.011 | **-5.458** | 0 | 546 | -0.151 | 0.143 | -1.060 | 0.290 | 268 | -0.021 | 0.006 | **-3.546** | 0 | 461 |
| Q21 AQ21 | -0.021 | 0.002 | **-8.326** | 0 | 160 | 0.988 | 0.488 | **2.027** | 0.047 | 55 | | | | | |
| Q22 AQ22 | -0.016 | 0.014 | -1.15 | 0.252 | 177 | 0.395 | 0.431 | 0.916 | 0.365 | 38 | | | | | |
| Q23 AQ23 | -0.060 | 0.008 | **-7.864** | 0 | 177 | -2.342 | 1.139 | **-2.057** | 0.044 | 56 | | | | | |
| Q24 AQ24 | 0.009 | 0.022 | 0.391 | 0.696 | 162 | 0.020 | 0.039 | 0.517 | 0.607 | 67 | | | | | |
| Y1 AY1 | -0.008 | 0.001 | **-6.061** | 0 | 1617 | 0.049 | 0.060 | 0.810 | 0.418 | 1218 | -0.003 | 0.005 | -0.718 | 0.473 | 1427 |
| **Panel B: Positive Trading Volume** | | | | | | | | | | | | | | | |
| Q11 AQ11 | -0.031 | 0.009 | **-3.546** | 0 | 233 | | | | | | | | | | |
| Q12 AQ12 | 0.028 | 0.042 | 0.651 | 0.516 | 213 | | | | | | | | | | |
| Q13 AQ13 | -0.006 | 0.016 | -0.373 | 0.71 | 241 | | | | | | | | | | |
| Q14 AQ14 | -0.037 | 0.020 | -1.855 | 0.065 | 181 | | | | | | | | | | |
| Q21 AQ21 | -0.028 | 0.002 | **-13.543** | 0 | 16 | | | | | | | | | | |
| Q22 AQ22 | -0.033 | 0.004 | **-9.434** | 0 | 19 | | | | | | | | | | |
| Q23 AQ23 | -0.059 | 0.011 | **-5.408** | 0.001 | 10 | | | | | | | | | | |
| Q24 AQ24 | -0.030 | 0.003 | **-8.487** | 0 | 8 | | | | | | | | | | |
| Y1 AY1 | -0.008 | 0.002 | **-6.413** | 0 | 1247 | -0.009 | 0.002 | **-4.38** | 0.001 | 115 | 0.003 | 0.002 | 2.104 | 0.038 | 109 |

## Table 7: Two-sample Variance comparison Test

This table reports variance ratio test between the returns of swaps (x) and the returns of replicating portfolios (y). $\sigma_x$ indicates the standard deviation for the first variable (swap returns) and $\sigma_y$ the standard deviation for the second variable (replicating portfolio returns). The test for $\sigma_x^2 = \sigma_y^2$ is given by the ratio of variances which is distributed as F with $N_{x-1}$ and $N_{y-1}$ degrees of freedom. The null hypothesis is that the ratio is equal to one. The three p-values are of three alternative hypotheses. The two-sided p-value is showed for Ha: ratio = 1. For the Ha: ratio<1 we show the lower one-sided p-value and for the Ha: ratio >1 the upper one-sided p-value.

| | Panel A : Full Sample | | | | | | | | Panel B: Positive Trading Volume | | | | | | | |
|---|---|---|---|---|---|---|---|---|---|---|---|---|---|---|---|---|
| | $\sigma_x$ | $\sigma_y$ | Ha: ratio=1 | Ha: ratio<1 | Ha: ratio>1 | F | $N_{x-1}$ | $N_{y-1}$ | $\sigma_x$ | $\sigma_y$ | Ha: ratio=1 | Ha: ratio<1 | Ha: ratio>1 | F | $N_{x-1}$ | $N_{y-1}$ |
| | | | | | | **German Market** | | | | | | | | | | |
| Q11 AQ11 | 0.016 | 0.017 | 0.273 | 0.136 | 0.864 | 0.905 | 483 | 483 | 0.017 | 0.020 | **0.004** | 0.002 | 0.998 | 0.696 | 475 | 155 |
| Q12 AQ12 | 0.017 | 0.018 | 0.096 | 0.048 | 0.952 | 0.860 | 487 | 487 | 0.017 | 0.017 | 0.502 | 0.251 | 0.749 | 0.915 | 463 | 135 |
| Q13 AQ13 | 0.010 | 0.011 | **0.002** | 0.001 | 0.999 | 0.772 | 563 | 563 | 0.010 | 0.011 | 0.075 | 0.038 | 0.962 | 0.795 | 546 | 140 |
| Q14 AQ14 | 0.014 | 0.014 | 0.550 | 0.275 | 0.725 | 0.950 | 535 | 535 | 0.014 | 0.016 | 0.110 | 0.055 | 0.945 | 0.792 | 517 | 103 |
| Q21 AQ21 | 0.017 | 0.017 | 0.999 | 0.500 | 0.500 | 1.000 | 151 | 151 | 0.017 | 0.060 | **0.000** | 0.000 | 1.000 | 0.083 | 139 | 3 |
| Q22 AQ22 | 0.010 | 0.010 | 0.998 | 0.501 | 0.499 | 1.000 | 166 | 166 | 0.011 | 0.010 | 0.955 | 0.478 | 0.522 | 1.200 | 132 | 3 |
| Q23 AQ23 | 0.009 | 0.009 | 0.993 | 0.497 | 0.503 | 0.999 | 167 | 167 | | | | | | | | |
| Q24 AQ24 | 0.010 | 0.011 | 0.337 | 0.168 | 0.832 | 0.855 | 151 | 151 | 0.011 | 0.004 | 0.650 | 0.675 | 0.325 | 5.654 | 118 | 1 |
| Y1 AY1 | 0.011 | 0.011 | 0.986 | 0.493 | 0.507 | 0.999 | 1588 | 1588 | 0.011 | 0.011 | 0.935 | 0.533 | 0.467 | 1.005 | 1572 | 1093 |
| | | | | | | **French Market** | | | | | | | | | | |
| Q11 AQ11 | 0.019 | 0.021 | 0.156 | 0.078 | 0.922 | 0.850 | 377 | 245 | | | | | | | | |
| Q12 AQ12 | 0.032 | 0.016 | **0.000** | 1.000 | 0.000 | 3.984 | 314 | 188 | | | | | | | | |
| Q13 AQ13 | 0.022 | 0.034 | **0.000** | 0.000 | 1.000 | 0.400 | 394 | 225 | | | | | | | | |
| Q14 AQ14 | 0.023 | 0.033 | **0.000** | 0.000 | 1.000 | 0.501 | 374 | 204 | | | | | | | | |
| Q21 AQ21 | 0.054 | 0.023 | **0.000** | 1.000 | 0.000 | 5.587 | 110 | 40 | | | | | | | | |
| Q22 AQ22 | 0.034 | 0.013 | **0.000** | 1.000 | 0.000 | 6.586 | 101 | 27 | | | | | | | | |
| Q23 AQ23 | 0.083 | 0.026 | **0.000** | 1.000 | 0.000 | 10.380 | 113 | 42 | | | | | | | | |
| Q24 AQ24 | 0.014 | 0.013 | 0.684 | 0.658 | 0.342 | 1.113 | 104 | 50 | | | | | | | | |
| Y1 AY1 | 0.010 | 0.011 | **0.027** | 0.014 | 0.986 | 0.870 | 1086 | 942 | 0.011 | 0.010 | 0.357 | 0.821 | 0.179 | 1.187 | 175 | 95 |
| | | | | | | **Spanish Market** | | | | | | | | | | |
| Q11 AQ11 | 0.0144 | 0.0171 | **0.0002** | 0.0001 | 0.9999 | 0.7048 | 473 | 461 | | | | | | | | |
| Q12 AQ12 | 0.0106 | 0.0109 | 0.5370 | 0.2685 | 0.7315 | 0.9432 | 451 | 439 | | | | | | | | |
| Q13 AQ13 | 0.0091 | 0.0093 | 0.6620 | 0.3310 | 0.6690 | 0.9611 | 493 | 477 | | | | | | | | |
| Q14 AQ14 | 0.0139 | 0.0142 | 0.5689 | 0.2844 | 0.7156 | 0.9479 | 461 | 445 | | | | | | | | |
| Y1 AY1 | 0.0082 | 0.0082 | 0.8272 | 0.4136 | 0.5864 | 0.9884 | 1410 | 1403 | 0.010 | 0.007 | **0.000** | 1.000 | 0.000 | 1.942 | 181 | 128 |

# Table 8: Trading Rule PD Test

This table reports the results from the PD trading rule. The Mean is the average daily profit, s.d. is its volatility (standard deviation) and we compute t-test under the null hypothesis that average daily profit is zero. P/L is the cumulative profit obtained by each strategy. Figures in boldface are significant at 1% confidence level. Transaction cost are trading fees (TF) and bid-ask spreads (B and A respectively, in percentage). For instance, in the case of Q11 and AQ11, costs are imputed as follows. Suppose the trader buys Q11 and sells AQ11. The respective (frictionless) prices are P(Q11) and P(AQ11). Including transaction cost implies that the price payed when buying Q11 is $(P(Q11)*(1+A_{Q11}))+TF$ and the price received by selling AQ11 is $1/3*(M1*(1-B_{M1}) + M2*(1-B_{M2}) + M3*(1-B_{M3})) - TF$.

| | Panel A: No Transaction costs | | | | | Panel B: Transaction costs | | | | |
|---|---|---|---|---|---|---|---|---|---|---|
| | N | Mean | s.d. | t-test | P/L | N | Mean | s.d. | t-test | P/L |
| **German Market** | | | | | | | | | | |
| **Q11** | 476 | 0.048 | 0.274 | **3.782** | 22.630 | 476 | -0.581 | 0.301 | **-42.194** | -276.684 |
| **Q12** | 474 | 0.054 | 0.254 | **4.600** | 25.470 | 474 | -0.641 | 0.298 | **-46.849** | -303.973 |
| **Q12** | 564 | 0.104 | 0.620 | **3.992** | 58.733 | 564 | -0.675 | 0.648 | **-24.748** | -380.811 |
| **Q14** | 534 | -0.022 | 0.593 | -0.843 | -11.560 | 534 | -0.809 | 0.618 | **-30.220** | -431.872 |
| **Q21** | 144 | 0.022 | 0.012 | **22.187** | 3.150 | 144 | -0.639 | 0.139 | **-55.109** | -91.959 |
| **Q22** | 167 | 0.029 | 0.013 | **29.159** | 4.867 | 167 | -0.747 | 0.184 | **-52.398** | -124.813 |
| **Q23** | 168 | 0.060 | 0.039 | **19.955** | 10.073 | 168 | -0.767 | 0.176 | **-56.571** | -128.779 |
| **Q24** | 152 | 0.021 | 0.211 | 1.208 | 3.147 | 152 | -0.625 | 0.229 | **-33.729** | -95.039 |
| **Y1** | 1396 | 0.014 | 0.013 | **41.090** | 19.302 | 1396 | -0.477 | 0.093 | **-191.030** | -665.580 |
| **French Market** | | | | | | | | | | |
| **Q11** | 306 | 0.094 | 0.555 | **2.960** | 28.733 | 306 | -1.890 | 0.682 | **-48.489** | -578.480 |
| **Q12** | 237 | 0.043 | 2.274 | 0.294 | 10.287 | 237 | -2.274 | 2.241 | **-15.619** | -538.830 |
| **Q12** | 281 | 0.360 | 2.111 | **2.861** | 101.263 | 281 | -2.698 | 2.224 | **-20.337** | -758.240 |
| **Q14** | 257 | -0.119 | 2.278 | -0.839 | -30.623 | 257 | -3.334 | 2.449 | **-21.821** | -856.800 |
| **Q21** | 39 | -0.077 | 3.089 | -0.156 | -3.003 | 39 | -1.984 | 3.124 | **-3.966** | -77.370 |
| **Q22** | 36 | -0.450 | 3.065 | -0.881 | -16.203 | 36 | -3.717 | 3.132 | **-7.121** | -133.800 |
| **Q23** | 51 | 2.601 | 7.418 | **2.504** | 132.663 | 51 | -1.123 | 7.656 | -1.048 | -57.290 |
| **Q24** | 62 | 0.018 | 0.112 | 1.301 | 1.147 | 62 | -2.212 | 0.200 | **-87.239** | -137.110 |
| **Y1** | 1102 | 0.074 | 1.126 | **2.195** | 82.042 | 1102 | -1.817 | 1.205 | **-50.068** | -2002.640 |
| **Spanish Market** | | | | | | | | | | |
| **Q11** | 400 | 0.014 | 0.047 | **6.186** | 5.783 | 400 | -2.579 | 0.465 | **-110.864** | -1031.468 |
| **Q12** | 402 | 0.016 | 0.100 | **3.249** | 6.523 | 402 | -2.987 | 0.445 | **-134.602** | -1200.691 |
| **Q13** | 414 | 0.040 | 0.203 | **3.978** | 16.437 | 414 | -2.913 | 0.567 | **-104.563** | -1206.135 |
| **Q14** | 417 | 0.031 | 0.039 | **16.214** | 12.880 | 417 | -3.001 | 0.426 | **-143.809** | -1251.406 |
| **Y1** | 1296 | 0.021 | 0.108 | **7.136** | 27.740 | 1296 | -2.307 | 0.309 | **-268.367** | -2949.257 |

## Table 9: Arbitrage Violations

This table reports the results from the AV trading rule. We present results for the full sample in Panel A, and in Panel B results for the restricted sample, including only days with positive trading volume. The column "No Trans. Cost" contains the percentage of violations of no-arbitrage conditions, assuming no transaction costs. The column "Sell Repl. Portf." contains the percentage of violations observed by selling the replicating portfolio and buying the basic contract, and including transaction costs. The column "Buy Repl. Portf." contains the percentage of violations observed by buying the replicating portfolio and selling the basic contract, and including transaction costs. Transaction cost are trading fees (TF) and bid-ask spreads (B and A respectively, in percentage). For instance in the case of Q11 and AQ11, costs are imputed as follows. Suppose the trader buys Q11 and sells AQ11. The respective (frictionless) prices are P (Q11) and P (AQ11). Including transaction cost implies that the price payed when buying Q11 is (P (Q11)*(1+AQ11)) +TF and the price received by selling AQ11 is 1/3*(M1*(1-BM1) + M2*(1-BM2) + M3*(1-BM3)) – TF.

|  | **German Market** | | | **French Market** | | | **Spanish Market** | | |
|---|---|---|---|---|---|---|---|---|---|
|  | No Trans. Cost | Sell Repl. Portf. | Buy Repl. Portf | No Trans. Cost | Sell Repl. Portf. | Buy Repl. Portf | No Trans. Cost | Sell Repl. Portf. | Buy Repl. Portf |
| **Panel A: Full Sample** | | | | | | | | | |
| **Q11 AQ11** | 93.29 | 0.81 | 0.00 | 92.55 | 0.00 | 0.62 | 64.71 | 0.00 | 0.21 |
| **Q12 AQ12** | 86.35 | 0.00 | 1.21 | 81.10 | 0.00 | 3.54 | 84.14 | 0.00 | 0.00 |
| **Q13 AQ13** | 98.95 | 0.00 | 0.87 | 100 | 0.00 | 0.00 | 76.11 | 0.00 | 0.00 |
| **Q14 AQ14** | 98.17 | 0.37 | 0.37 | 94.8 | 2.60 | 1.12 | 80.26 | 0.00 | 0.00 |
| **Q21 AQ21** | 90.63 | 0.00 | 0.00 | 62.29 | 0.00 | 7.14 |  |  |  |
| **Q22 AQ22** | 98.87 | 0.00 | 0.57 | 92.31 | 0.00 | 2.56 |  |  |  |
| **Q23 AQ23** | 100.00 | 0.00 | 0.00 | 93.33 | 0.00 | 0.00 |  |  |  |
| **Q24 AQ24** | 98.15 | 0.00 | 1.85 | 97.06 | 0.00 | 1.47 |  |  |  |
| **Y1 AY1** | 63.39 | 0.06 | 0.06 | 85.23 | 0.00 | 0.49 | 28.99 | 0.07 | 0.00 |
| **Panel B: Positive Trading Volume** | | | | | | | | | |
| **Q11 AQ11** | 90.13 | 0.86 | 0.00 |  |  |  | 33.33 | 0.00 | 0.00 |
| **Q12 AQ12** | 84.04 | 0.00 | 0.94 | 100.00 | 0.00 | 0.00 | 100.00 | 0.00 | 0.00 |
| **Q13 AQ13** | 98.74 | 0.00 | 1.26 | 100.00 | 0.00 | 0.00 | 66.67 | 0.00 | 0.00 |
| **Q14 AQ14** | 97.79 | 0.00 | 0.55 |  |  |  | 100.00 | 0.00 | 0.00 |
| **Q21 AQ21** | 93.33 | 0.00 | 0.00 |  |  |  |  |  |  |
| **Q22 AQ22** | 100.00 | 0.00 | 0.00 |  |  |  |  |  |  |
| **Q23 AQ23** | 100.00 | 0.00 | 0.00 |  |  |  |  |  |  |
| **Q24 AQ24** | 100.00 | 0.00 | 0.00 |  |  |  |  |  |  |
| **Y1 AY1** | 59.79 | 0.08 | 0.00 | 82.6 | 0.00 | 0.00 | 26.61 | 0.00 | 0.00 |